\documentclass[10pt]{article}
\RequirePackage{titlesec}
\RequirePackage{amsmath,esdiff}
\RequirePackage{amssymb}
\RequirePackage{authblk}
\DeclareMathOperator{\csch}{cosech}

\RequirePackage[normalem]{ulem}
\RequirePackage{bbold,bm}
\RequirePackage{BOONDOX-cal}
\RequirePackage[colorlinks=true,linkcolor=blue,citecolor=red,urlcolor=red]{hyperref}
\RequirePackage{geometry}
\RequirePackage{setspace}
\RequirePackage{cite}
\RequirePackage{flushend}
\RequirePackage{mathrsfs}
\RequirePackage{hyperref}
\RequirePackage{graphicx}
\RequirePackage{cancel}
\RequirePackage{array,esint}
\RequirePackage[most]{tcolorbox}
\graphicspath{}
\titleformat{\section}{\fontsize{12}{12}\bfseries}{\thesection}{1em}{}
\begin{document}
\title{\textbf{Near horizon approximation and beyond for a two-level atom falling into a Kerr-Newman black hole}}
\author{\textbf{Soham Sen${}^{a*}$, Rituparna Mandal${}^{a\dagger}$ and Sunandan Gangopadhyay${}^{a\ddagger}$}}
\affil{{${}^a$ Department of Astrophysics and High Energy Physics}\\
{S.N. Bose National Centre for Basic Sciences}\\
{JD Block, Sector III, Salt Lake, Kolkata 700 106, India}}
\date{}
\maketitle
\begin{abstract}
\noindent  In this work we investigate the phenomena of acceleration radiation for a two-level atom falling into the event horizon of a Kerr-Newman black hole. In (\href{https://link.aps.org/doi/10.1103/PhysRevD.104.065006}{Phys. Rev. D 104 (2021) 065006}), it has been shown that conformal quantum mechanics has a connection to the generated Planck-like spectrum due to acceleration radiation. In this particular aspect, the near-horizon approximation played a significant role. In (\href{https://link.aps.org/doi/10.1103/PhysRevD.106.025004}{Phys. Rev. D 106 (2022) 025004}), we have used the beyond near horizon approximation to show that the excitation probability attains a Planck-like spectrum irrespective of the non-existence of an underlying conformal symmetry for a general class of static spherically symmetric black holes. In our current analysis, we have gone beyond the near horizon approximation for the rotating and charged case and without the consideration of the conformal symmetry, we no longer observe an overall Planck-like spectrum. We instead observe that the excitation probability consists of several lower incomplete gamma functions which deform the near-horizon conformal behaviour of the spectrum. Finally, we have computed the von-Neumann entropy which is also known as the horizon brightened acceleration radiation entropy or the HBAR entropy. The calculation of the rate of change of the von Neumann entropy suggests that there is a loss of thermality because of the inclusion of the contributions from the beyond-near horizon terms.
\end{abstract}
\section{Introduction}
\footnote{{}\\
{$*$sensohomhary@gmail.com, soham.sen@bose.res.in}\\
{$\dagger$drimit.ritu@gmail.com}\\
{$\ddagger$sunandan.gangopadhyay@gmail.com}}
\noindent The greatest endeavour of modern theoretical physics is to unify multiple mainstream descriptions of the underlying theories of nature under a more fundamental theoretical description. General relativity\cite{Einstein15,Einstein16} and quantum mechanics, being the two most intriguing theoretical descriptions of the universe, are the most complicated ones to unify in a single theory. The theoretical framework needed to write down a quantum theory of gravity is quantum field theory. The quantum field theoretic description of gravity can be considered a successful effective field theory. The energy scale that is presently accessible to us is well described by general relativity and there is a theory of quantum gravity that works well at ordinary energy scales. There are still some issues present when writing down a quantum theory of gravity at very high energies. The unsolved issue is about the ultra-violet completion of gravity. 
%Although there have been several successful attempts to unify other fundamental theories like quantum field theory, electrodynamics, and the weak and the strong nuclear forces, none have been able to unify quantum mechanics and general relativity successfully. 
The efforts to unify gravity, thermodynamics, and geometry led to several ground-breaking discoveries like black hole thermodynamics\cite{Hawking,Hawking2,Hawking3,Bekenstein,Bekenstein2}, Hawking radiation\cite{Hawking,Hawking2,Hawking3,SG0a,SG0b,SG0c,SG0d,SG0e,SG0f}, particle emission from a black hole\cite{Page, Page2,Page3}, acceleration radiation\cite{Fulling21,Davies,DeWitt,Unruh2,Muller,Vanzella,Higuichi,Conformal1, Conformal2,Fulling2,Fulling,Fulling0,Ordonez1,Ordonez2,Ordonez3,
Ordonez4,SG, OTM, OTM2,OTM3}, and the Unruh effect\cite{Unruh}. Hawking radiation is observed from a black hole if one considers quantum effects in curved space-time. Another phenomenon describing the spontaneous creation of particles is the Schwinger mechanism\cite{Schwinger}. In this case, a very strong electric field is applied which results in the production of pairs of particles and antiparticles.  In the case of the Unruh-Fulling effect, it has been observed that if an observer is uniformly accelerated (with acceleration `$a$') through an inertial vacuum, the observer can detect a thermal bath of particles (quanta of the fields) with  a temperature
\begin{equation}\label{1.0}
T=\frac{\hbar a}{2\pi k_B c}
\end{equation}
where $k_B$ denotes the Boltzmann constant. In contrast to the accelerated observer, the inertial observer does not detect any such swarm of excited particles. In this process, the two-level atom jumps to the excited state with the simultaneous emission of a virtual photon provided that it was previously prepared in the ground state. By interrupting this virtual process one can observe real photons. It can be interpreted in the following way. The atom accelerates away thereby  converting the virtual photon into a real photon. The energy used in this procedure is provided by the external force field which is driving the center of mass motion of the two-level atom. The same particle creation scenario from a different perspective was observed in \cite{Fulling2} for a two-level atom freely falling into the event horizon of a Schwarzschild black hole. In this process, the two-level atom emits real photons and the excitation probability is similar to that of the excitation probability for the case of a mirror accelerating with respect to a fixed two-level atom in flat space-time. It was also observed that in this process the freely falling atom feels the effect of thermal radiation similar to that of Hawking radiation. The von-Neumann entropy considering this radiation was termed as the horizon brightened acceleration radiation entropy or the HBAR entropy \cite{Fulling2}. Very few works also considered the effects of the conformal symmetry as a contributing factor towards near horizon analysis of a black hole \cite{Ordonez1,Ordonez2,Ordonez3,Ordonez4,Conformal1,Conformal2}. In \cite{Ordonez1}, it was observed that a near horizon analysis, in principle, harbours a conformal symmetry leading to a Planck-like distribution for the simultaneously emitted photons from two-level atoms freely falling into static spherically symmetric black holes. Later in \cite{Ordonez2}, the same analysis was carried out for a Kerr black hole, and a similar Planck factor was observed if viewed from the perspective of a co-rotating observer near the event horizon of the rotating black hole. The dominant cause for the Planck-like distribution was considered to be the underlying conformal symmetry arising due to near horizon approximation. Recently, in \cite{OTM2}, we considered two-level atoms freely falling into a general class of static spherically symmetric black holes with the simultaneous emission of scalar photons. Our analysis revolved around the consideration of the ``beyond near horizon analysis" which in principle breaks the conformal symmetry present in the earlier analysis. We found that apart from some changes in the coefficient, the Planck-like factor retained its integrity concluding that the conformal symmetry does not play a crucial role in the case of the existence of a Planck-like factor in the analytical form of the excitation probabilities. 

\noindent Inspired by the existing works in these fields,  we consider a two-level uncharged atom freely falling into the event horizon of a Kerr-Newman black hole with a perfectly reflecting mirror at the event horizon of the black hole in this work. We have also implemented the ``beyond near horizon" approximation in our current analysis. Our aim is also to investigate whether one can obtain the Planck-like distribution (or it gets deformed) as observed for a Kerr black hole \cite{Ordonez2} in the case of a beyond-near horizon analysis. It is also important to note that going beyond the near horizon analysis for a black hole without spherical symmetry is important in its own right as the rotating black holes are the only observed ones in the universe. Also, investigation beyond the near horizon for rotating black holes has been missing in the literature. In this respect, this is the first work in this direction. Taking into account the effect of rotation not only makes the investigation more complete but also more challenging. We have focussed on the differences between our analytical results for the transition probability with those that have been in the literature \cite{Ordonez2}. In our analysis the effect of the curvature of rotating spacetime felt by the two-level atom becomes significant and as a consequence of this, we no longer observe a blackbody-like emission spectrum. The spectrum found through our calculation is a deformed one.

\noindent The organization of the paper goes as follows. In section(\ref{Section2}), we compute the scalar field solution from the covariant Klein-Gordon equation in the Kerr-Newman background. In section(\ref{Section3}), we write down the trajectories of the freely falling atom in the event horizon of the Kerr-Newman black hole. In section(\ref{Section4}), we compute the excitation and absorption probabilities. In section(\ref{Section_Plot}), we have given some plots depicting the deformed nature of the transition probability obtained in this work from the usual Planck-like spectrum obtained in the case of the near horizon approximation. In section(\ref{Section6}), we calculate the horizon brightened acceleration radiation (HBAR) entropy for such a process and then finally conclude this work.  
\section{The Kerr Newman geometry and the beyond near horizon approximation}\label{Section2}
In this section, we shall compute the solution of the Klein-Gordon equation in the Kerr-Newman black hole background. 
The exact solution of Einstein's field equations for a charged and rotating black hole of mass $M$, charge $Q$, and angular momentum $J$ in the Boyer-Lindquist coordinates ($\{t,r,\theta,\phi\}$) is given by \cite{Kerr,Kerr_Newman,Boyer_Lindquist}
\begin{equation}\label{1.1}
\begin{split}
ds^2=&-\frac{\Delta_Q}{\rho^2}(dt-a\sin^2\theta d\phi)^2+\frac{\rho^2}{\Delta_Q}dr^2+\rho^2d\theta^2+\frac{\sin^2\theta}{\rho^2}[(r^2+a^2)d\phi-adt]^2
\end{split}
\end{equation}
where $a=\frac{J}{M}$, $\Delta_Q=r^2-2Mr+a^2+Q^2$, and $\rho^2=r^2+a^2\cos^2\theta$. In this case, we are considering geometrized units ($c=G=1$). By setting $\Delta_Q=0$, one can obtain the outer and inner horizons of the Kerr-Newman black hole as $r_\pm=M\pm\sqrt{M^2-a^2-Q^2}$. The line element in eq.(\ref{1.1}), can be recast in the following form
\begin{equation}\label{1.2}
ds^2=-\frac{\rho^2\Delta_Q}{\Sigma^2}dt^2+\frac{\rho^2}{\Delta_Q}dr^2+\rho^2d\theta^2+\frac{\Sigma^2}{\rho^2}\sin^2\theta(d\phi-\varpi dt)^2
\end{equation}
where $\Sigma^2=(r^2+a^2)^2-\Delta_Q a^2\sin^2\theta$ and $\varpi=-\frac{g_{t\phi}}{g_\phi\phi}=-\frac{a(Q^2-2Mr)}{(r^2+a^2)^2-\Delta_Qa^2\sin^2\theta}$~. The position-dependent angular velocity ($\varpi$) with respect to an external reference frame can be reduced to the angular velocity ($\tilde{\Omega}$) of the black hole while approaching the external horizon of the Kerr-Newman black hole. The form $\tilde{\Omega}$ reads
\begin{equation}\label{1.3}
\tilde{\Omega}=\lim\limits_{r\rightarrow r_+}\varpi=\frac{a}{r_+^2+a^2}
\end{equation}
where we have made use of the identity $r_+^2+a^2=2Mr_+-Q^2$. To proceed further, we need to compute the determinant and the inverse components of the metric tensor for the Kerr-Newman geometry. The determinant $g$ of the metric tensor reads
\begin{equation}\label{1.4}
g=g_{tt}g_{\phi\phi}-g_{t\phi}^2=-\rho^4\sin^2(\theta)~,
\end{equation}
and the non-vanishing metric components of $g^{\mu\nu}$ reads
\begin{equation}\label{1.5}
\begin{split}
g^{tt}&=-\frac{(r^2+a^2)^2-\Delta_Qa^2 \sin^2\theta}{\rho^2\Delta_Q^2 },~g^{rr}=\frac{\Delta_Q}{\rho^2},~
g^{\theta\theta}=\frac{1}{\rho^2},~g^{t\phi}=g^{\phi t}=\frac{a(\Delta_Q-(r^2+a^2))}{\rho^2\Delta_Q},~g^{\phi\phi}=\frac{\Delta_Q-a^2\sin^2\theta}{\rho^2\Delta_Q\sin^2\theta}~.
\end{split}
\end{equation}
In our analysis, we are considering the case of acceleration radiation of a scalar field of mass $\sigma_0$ from an atom of mass $\sigma_a$ near the outer horizon of a Kerr-Newman black hole. The covariant Klein-Gordon equation for a scalar field $\Psi$ reads
\begin{equation}\label{1.6}
\frac{1}{\sqrt{-g}}\partial_{\mu}(\sqrt{-g}g^{\mu\nu}\partial_\nu\Psi)-\sigma_0^2\Psi=0~.
\end{equation}
Using the form of $g$ from eq.(\ref{1.4}) and the non-vanishing components of the inverse of the metric tensor from eq.(\ref{1.5}), we can recast the above equation in the following form
\begin{equation}\label{1.7}
\begin{split}
&-\frac{\Sigma^2}{\Delta_Q}\frac{\partial^2\Psi}{\partial t^2}+\frac{2a(Q^2-2 M r)}{\Delta_Q}\frac{\partial^2\Psi}{\partial t\partial \phi}+\frac{\partial}{\partial r}\left(\Delta_Q\frac{\partial \Psi}{\partial r}\right)+\frac{1}{\sin\theta}\frac{\partial}{\partial \theta}\left(\sin\theta\frac{\partial \Psi}{\partial\theta}\right)+\left(\frac{1}{\sin^2\theta}-\frac{a^2}{\Delta_Q}\right)\frac{\partial^2\Psi}{\partial\phi^2}=\sigma_0^2\rho^2\Psi.
\end{split}
\end{equation}
In order to solve eq.(\ref{1.7}), we denote $\Psi=\psi_{\nu l m}$ in terms of the frequency ($\nu$) and the quantum numbers $l,m$. We now consider a separation of variables for $\psi_{\nu l m}$ as follows
\begin{equation}\label{1.8}
\psi_{\nu l m}(t,r,\theta,\phi)=e^{-i\nu t}R_{\nu l m}(r)\Theta_{\nu l m}(\theta)e^{im\phi}
\end{equation}
where we will use $R(r)$ and $\Theta(\theta)$ instead of $R_{\nu lm}(r)$ and $\Theta_{\nu lm}(\theta)$ respectively throughout our analysis. We can now define a new frame corotating with the black hole with an angular velocity $\tilde{\Omega}$ by introducing the following coordinate changes
\begin{equation}\label{1.9}
\tilde{t}=t~,~\tilde{\phi}=\phi-\tilde{\Omega} t~,
\end{equation}
and the form of the shifted frequency is given as follows
\begin{equation}\label{1.10}
\tilde{\nu}=\nu-m\tilde{\Omega}=\nu-\frac{am}{r_+^2+a^2}~.
\end{equation}
Using the new quantities in eq.(s)(\ref{1.9},\ref{1.10}), we can rewrite the form of the scalar field in eq.(\ref{1.8}), as
\begin{equation}\label{1.11}
\psi_{\tilde{\nu}lm}(t,r,\theta,\tilde{\phi})=e^{-i\tilde{\nu}t}R(r)\Theta(\theta)e^{im\tilde{\phi}}~.
\end{equation}
Using eq.(\ref{1.8}) back in eq.(\ref{1.7}), we can separate the equations involving the radial part of the solution and the $\theta$ dependent part of the solution. The radial equation reads
\begin{equation}\label{1.12}
\begin{split}
&\frac{d}{dr}\left[\Delta_Q\frac{dR}{dr}\right]+\biggr[\frac{(r^2+a^2)^2\nu^2+2a\nu m(Q^2-2Mr)+a^2m^2}{\Delta_Q}-a^2\nu^2-\sigma_0^2r^2-\beta \biggr]R=0
\end{split}
\end{equation}
with $\beta$ being the separation constant and the differential equation involving $\Theta$ reads
\begin{equation}\label{1.13}
\begin{split}
&\frac{1}{\sin\theta}\frac{d}{d\theta}\left[\sin\theta\frac{d\Theta}{d\theta}\right]+\biggr[a^2\nu^2\cos^2\theta-\frac{m^2}{\sin^2\theta}+\beta-\sigma_0^2a^2\cos^2\theta\biggr]\Theta=0~.
\end{split}
\end{equation}
In order to solve the radial equation, we use the beyond-near-horizon approximation. The form of $\Delta_Q$ and its higher order derivatives in the beyond near horizon approximation takes the following form
\begin{align}
\Delta_Q(r)&\simeq(r-r_+)\Delta_Q'(r_+)+\frac{1}{2}(r-r_+)^2\Delta_Q''(r_+)=y\Delta'_{Q+}+\frac{1}{2}y^2\Delta''_{Q+}\label{1.14}~,\\
\Delta_Q'(r)&\simeq\Delta'_{Q+}+y\Delta''_{Q+}\label{1.15}~,\\
\Delta_Q''(r)&\simeq\Delta''_{Q+}=2\label{1.16}
\end{align}
where $\Delta_Q'(r_+)=\Delta'_{Q+}$, $\Delta_Q''(r_+)=\Delta''_{Q+}$, $r-r_+=y$, and $y$ is much smaller with respect to the outer horizon radius.
When sufficiently close to the outer horizon radius the Killing vector is time-like and one can define the associated surface gravity as 
\begin{equation}\label{1.17}
\begin{split}
\kappa&=-\frac{1}{2}(\nabla_\mu\xi_\nu)(\nabla^\mu\xi^\nu)=\frac{r_+-r_-}{2(r_+^2+a^2)}=\frac{\Delta'_{Q+}}{2(r_+^2+a^2)}~.
\end{split}
\end{equation}
Using the beyond near horizon approximation, we can recast the radial equation in eq.(\ref{1.12}) as
\begin{equation}\label{1.18}
\begin{split}
&\frac{1}{y}\left[\frac{d}{dx}\left[y\left(1+\frac{y}{\Delta'_{Q+}}\right)\frac{dR(y)}{dy}\right]\right]+\biggr[\frac{\tilde{\nu}^2}{4\kappa^2}\frac{1}{y^2+}+\biggr[4\nu\biggr[\frac{\tilde{\nu}r_+}{2\kappa}+\frac{1}{2}am\biggr]-\frac{\tilde{\nu}^2}{4\kappa^2}-(a^2\nu^2+\sigma_0^2r_+^2+\beta)\biggr]\frac{1}{\Delta'_{Q+} y}\biggr]R(y)=0~.
\end{split}
\end{equation}
In order to solve the above radial equation, we consider the following relation
\begin{equation}\label{1.19}
R(y)=\left(y+\frac{y^2}{\Delta'_{Q+}}\right)^{-\frac{1}{2}}u(y)~.
\end{equation}
Substituting the above equation back in eq.(\ref{1.18}), we can recast the radial equation in terms of $u(y)$ as follows
\begin{equation}\label{1.20}
y^2u''(y)+\left[\left(\frac{\tilde{\nu}}{2\kappa}\right)^2+\frac{1}{4}\right]u(y)-\mathcal{B}yu(y)=0
\end{equation}
where  $'$ denotes differentiation with respect to the variable $y$ and
\begin{equation}\label{1.21}
\begin{split}
\mathcal{B}&=\frac{2}{\Delta'_{Q+}}\left[\left(\frac{\tilde{\nu}}{2\kappa}\right)^2+\frac{1}{4}\right]+\frac{a^2\nu^2+\beta+\sigma_0^2r_+^2}{\Delta'_{Q+}}-\frac{4\nu}{\Delta'_{Q+}}\left(\frac{\tilde{\nu}}{2\kappa}r_++\frac{1}{2}am\right)~.
\end{split}
\end{equation}
It is very important to observe that the $\mathcal{B}$ dependent linear term in eq.(\ref{1.20}) arises due to the consideration of the beyond near horizon approximation only. If we now set $\mathcal{B}=0$, we obtain the usual scaling invariant equation \cite{Ordonez1,Ordonez2} exhibiting the asymptotic conformal symmetry. To solve eq.(\ref{1.21}), we consider a trial solution of the form
\begin{equation}\label{1.22}
u(y)=\sum\limits_{k=0}^\infty\mathcal{A}_ky^{p+k} 
\end{equation}
with $p$ being a constant.  Substituting the above equation back in eq.(\ref{1.20}), we obtain the following relation
\begin{equation}\label{1.23}
\sum\limits_{k=0}^\infty\mathcal{A}_ky^{p+k}\left[(p+k)(p+k-1)-\mathcal{B}y+\frac{\tilde{\nu}^2}{4\kappa^2}+\frac{1}{4}\right]=0.
\end{equation}
In order to compute the individual unknown coefficients $\mathcal{A}_k$ for each $k$, we shall compare each equation containing different powers of the variable $y$ from eq.(\ref{1.23}) to zero. The equation from the $p^{\text{th}}$ power of order $y$ as follows
\begin{equation}\label{1.24}
p^2-p+\frac{\tilde{\nu}^2}{4\kappa^2}+\frac{1}{4}=0~.
\end{equation} 
The solution of eq.(\ref{1.24}) is given as follows
\begin{equation}\label{1.25}
p=\frac{1}{2}\pm i\frac{\tilde{\nu}}{2\kappa}~.
\end{equation}
From the higher order equations in $y$, one can obtain the recursion relation involving the coefficients $\mathcal{A}_k$. Hence, we can represent every coefficient $\mathcal{A}_k$ in terms of the coefficient $\mathcal{A}_0$ using the recursion relation as follows
\begin{equation}\label{1.26}
\mathcal{A}_k=\frac{\mathcal{B}^k\mathcal{A}_0\Gamma\left[1+i\frac{\tilde{\nu}}{\kappa}\right]}{k!\Gamma\left[k+1+i\frac{\tilde{\nu}}{\kappa}\right]}~.
\end{equation}
As $\mathcal{A}_0$ is an arbitrary constant, we set its value equal to $1$. Using the form of the coefficients in the above equation, we obtain the formal solution of eq.(\ref{1.20}) given as
\begin{equation}\label{1.27}
u(y)=y^{\frac{1}{2}+\frac{i\tilde{\nu}}{2\kappa}}\sum\limits_{k=0}^\infty\frac{y^k\mathcal{B}^k\Gamma\left[1+\frac{i\tilde{\nu}}{\kappa}\right]}{k!\Gamma\left[k+1+\frac{i\tilde{\nu}}{\kappa}\right]}~.
\end{equation}
As we are considering a very small value of the variable $y$, we shall keep up to the linear order term in $y$ in the above equation. Using eq.(\ref{1.27}) back in eq.(\ref{1.19}), we obtain the form of $R(y)$ upto $\mathcal{O}(y)$ to be 
\begin{equation}\label{1.28}
\begin{split}
R(y)=&\left(y+\frac{y^2\Delta''_{Q+}}{2\Delta'_{Q+}}\right)^{-\frac{1}{2}}y^{\frac{1}{2}+\frac{i\tilde{\nu}}{2\kappa}}\sum\limits_{k=0}^\infty\frac{y^k\mathcal{B}^k\Gamma\left[1+\frac{i\tilde{\nu}}{\kappa}\right]}{k!\Gamma\left[k+1+\frac{i\tilde{\nu}}{\kappa}\right]}\\
\simeq&y^{-\frac{i\tilde{\nu}}{2\kappa}}\left(1+\left(\frac{\kappa^2\mathcal{B}^2}{\kappa^2+\tilde{\nu}^2}-\frac{1}{2\Delta'_{Q+}}\right)y+\frac{i\tilde{\nu}\kappa\mathcal{B}}{\kappa^2+\tilde{\nu}^2}y\right)~.
\end{split}
\end{equation}
With the radial part of the solution in hand, we shall now proceed to obtain the $\theta$ part of the solution. We consider that the angle $\theta$ will attain a value $\theta_+$ when the particle reaches near the outer horizon radius $r_+$. To proceed further, we define a new variable as follows
\begin{equation}\label{1.29}
\varphi=\theta-\theta_+
\end{equation}
where $\varphi\ll\theta_+$. We can now Taylor expand any function $f(\theta)$ about $\theta_+$ up to $\mathcal{O}(\varphi)$ as $f(\theta)\simeq f(\theta_+)+\varphi f'(\theta_+)$. Up to $\mathcal{O}(\varphi)$, we can recast eq.(\ref{1.13}) as
\begin{equation}\label{1.30}
\begin{split}
&\frac{d}{d\varphi}\left[[1+\varphi\cot\theta_+]\frac{d\Theta(\varphi)}{d\varphi}\right]+\biggr[a^2\cos^2\theta_+[\nu^2-\sigma_0^2]-\frac{m^2}{\sin^2\theta_+}+\beta\biggr]\Theta(\varphi)\\&+\varphi\biggr[a^2\cos^2\theta_+(cot\theta_+-2\tan\theta_+)(\nu^2-\sigma_0^2)+\beta\cot\theta_++\frac{m^2\cot\theta_+}{\sin^2\theta_+}\biggr]\Theta(\varphi)=0~.
\end{split}
\end{equation}
In order to solve the above equation we consider the following form of $\Theta(\varphi)$
\begin{equation}\label{1.31}
\Theta(\varphi)=(1+\varphi \cot\theta_+)^{-\frac{1}{2}}\mathcal{S}(\varphi)~.
\end{equation}
By using the above relation, we can recast eq.(\ref{1.30}) in the following form
\begin{equation}\label{1.32}
\frac{d^2\mathcal{S}(\varphi)}{d\varphi^2}+\mathcal{p}_1\mathcal{S}(\varphi)-\mathcal{p}_2\varphi\mathcal{S}(\varphi)=0
\end{equation}
where the analytical forms of $\mathcal{p}_1$ and $\mathcal{p}_2$ are given as follows
\begin{align}
\mathcal{p}_1&=\frac{\cot^2\theta_+}{4}+a^2\cos^2\theta_+(\nu^2-\sigma_0^2)+\beta-\frac{m^2}{\sin^2\theta_+}~,\label{1.33}\\
\mathcal{p}_2&=a^2\sin2\theta_+(\nu^2-\sigma_0^2)-\frac{2m^2\cot\theta_+}{\sin^2\theta_+}+\frac{\cot^3\theta_+}{4}~.\label{1.34}
\end{align}
Now eq.(\ref{1.32}) has an exact solution of the form 
\begin{equation}\label{1.35}
\mathcal{S}(\varphi)=\mathcal{C}_1\text{Ai}\left(\frac{-\mathcal{p}_1+\varphi \mathcal{p}_2}{\mathcal{p}_2^{\frac{2}{3}}}\right)+\mathcal{C}_2\text{Bi}\left(\frac{-\mathcal{p}_1+\varphi \mathcal{p}_2}{\mathcal{p}_2^{\frac{2}{3}}}\right)
\end{equation}
with $\text{Ai}$ and $\text{Bi}$ denoting the `AiryAi' and `AiryBi' functions respectively, and $\mathcal{C}_1$ and $\mathcal{C}_2$ being the integration constants. When the particle approaches the event horizon of the Kerr-Newman black hole, according to our earlier consideration, $\theta\rightarrow\theta_+$ and $\varphi\rightarrow 0$, and therefore we do not explicitly write the form of the solution up to $\mathcal{O}(\varphi)$. Therefore, in this limit, eq.(\ref{1.31}) reduces to the following form
\begin{equation}\label{1.36}
\lim\limits_{\varphi\rightarrow 0}\Theta(\varphi)=\Theta_+=\mathcal{C}_1\text{Ai}\left(-\frac{\mathcal{p}_1}{\mathcal{p}_2^{\frac{2}{3}}}\right)+\mathcal{C}_2\text{Bi}\left(\frac{-\mathcal{p}_1}{\mathcal{p}_2^{\frac{2}{3}}}\right)~.
\end{equation}
We can now rewrite the solution of the Klein-Gordon equation as follows
\begin{equation}\label{1.37}
\begin{split}
\psi_{\tilde{\nu}lm}&=e^{-i\tilde{\nu}t}y^{-\frac{i\tilde{\nu}}{2\kappa}}\biggr[1+\biggr[\frac{\kappa^2\mathcal{B}^2}{\kappa^2+\tilde{\nu}^2}-\frac{1}{2\Delta'_{Q+}}\biggr]y+\frac{i\tilde{\nu}\kappa\mathcal{B}y}{\kappa^2+\tilde{\nu}^2}\biggr]\Theta_+e^{im\tilde{\phi}}~.
\end{split}
\end{equation}
Before proceeding further, we need to define the Boulware vacuum in our current analysis. Due to the non-existence of the conformal symmetry in the mode solution given in eq.(\ref{1.37}) because of the presence of terms of $\mathcal{O}(y)$, we can no longer use the same argument presented in Appendix B of \cite{Ordonez2}. Another way to deal with this problem was proposed in \cite{QuantumVacuum}. This was to replace the past and future null-infinities ($\mathcal{I}^{\pm}$) with Dirichlet-like bounded-domain boundary conditions \cite{Ordonez2,QuantumVacuum}. Following the argument presented in \cite{QuantumVacuum}, one can consider a vacuum state that is corotating with its detector. The physical reason behind the assumption in \cite{QuantumVacuum} is that the rotating Killing vectors become spacelike outside the light cylinder. However, in our present analysis, the associated rotating Killing vector is still time-like within the ergosphere of the black hole. Hence, in order to properly define the Boulware-like vacuum states, we need to consider an anti de-Sitter like construction where the Kerr-Newman black hole is kept inside a box of some radius $r$ from the centre with perfectly reflecting boundaries. This set up can be interpreted as placing a mirror in our analysis.%Now one can consider a perfectly reflecting mirror defining the boundary (corotating with the detector) and the boundary is at such a distance $r$ (from the $r=0$ point) that any point on the mirror does not exceed the speed of light at an instant. This surface is also known as the ``\textit{Speed of Light}" surface or the SOL surface. Following the calculation in \cite{QuantumVacuum}, one can conclude that if the boundary (mirror) is placed within the SOL surface, the particle detector corotating with the rotating vacuum state does not detect any presence of quanta. If the boundary is placed outside this SOL surface, the problems regarding the existence of superradiant modes \cite{SuperRadiant,SuperRadiant2} and Unruh-Starobinskii effect rebounds \cite{Starobinskii} resulting in the ambiguity in defining a suitable vacuum state. Hence, if one puts the mirror (creating the boundary condition) inside the SOL surface, one can uniquely define the Boulware vacuum $|B\rangle$.

\noindent In order to compute the excitation probability we need to obtain the equations governing the trajectory of the two-level atom freely falling into the event horizon of the black hole.
\section{Trajectories of the atom freely falling into the black hole}\label{Section3}
In terms of the time-like and azimuthal Killing vectors $\bm{\vartheta}$ and $\bm{\varsigma}$, we can define two conserved quantities as follows
\begin{align}
E&=-\bm{\vartheta}.\bm{p}=-g_{tt}\frac{dt}{d\lambda}-g_{t\phi}\frac{d\phi}{d\lambda}~,\label{1.38}\\
L_z&=\bm{\varsigma}.\bm{p}=g_{t\phi}\frac{dt}{d\lambda}+g_{\phi\phi}\frac{d\phi}{d\lambda}\label{1.39}
\end{align}
where $E$ denotes the energy, $L_z$ denotes the angular momentum, $\bm{p}$ denotes the four momentum of the atom, and $\lambda$ denotes the affine parameter. One can make an alternative choice of the affine parameter by denoting $\tau=\lambda\sigma_a$ ($\sigma_a$ is the mass of the two-level atom) and define the new quantities as follows
\begin{align}
\mathcal{e}=\frac{E}{\sigma_a}=-g_{tt}\frac{dt}{d\tau}-g_{t\phi}\frac{d\phi}{d\tau}~,~~
\mathcal{l}=\frac{L_z}{\sigma_a}=g_{t\phi}\frac{dt}{d\tau}+g_{\phi\phi}\frac{d\phi}{d\tau}\label{1.40}
\end{align} 
where $\mathcal{e}$ denotes the energy per unit mass and $\mathcal{l}$ denotes the angular momentum per unit mass. Using the above two relations, one can obtain the following two trajectory equations
\begin{align}
\rho^2\frac{dt}{d\tau}&=a(\mathcal{l}-a\mathcal{e}\sin^2\theta)+\frac{r^2+a^2}{\Delta_Q}\mathcal{K}(r)~,\label{1.41}\\
\rho^2\frac{d\phi}{d\tau}&=-\left(a\mathcal{e}-\frac{\mathcal{l}}{\sin^2\theta}\right)+\frac{a}{\Delta_Q}\mathcal{K}(r)\label{1.42}
\end{align}
where $\mathcal{K}(r)=(r^2+a^2)\mathcal{e}-a\mathcal{l}$. One can also define one other conserved quantity as follows
\begin{equation}\label{1.43}
\mathcal{q}=\frac{\mathcal{Q}}{\sigma_0^2}=\left(\frac{p_\theta}{\sigma_0}\right)^2+\cos^2\theta\left[a^2(1-\mathcal{e}^2)+\left(\frac{\mathcal{l}}{\sin\theta}\right)^2\right]~.
\end{equation}
Following the standard procedure, one can obtain the following two trajectory equations as well
\begin{align}
\rho^2\frac{dr}{d\tau}&=-\sqrt{\mathcal{R}(r)}~,\label{1.44}\\
\rho^2\frac{d\phi}{d\tau}&=\pm\sqrt{\mathcal{U}(\theta)}\label{1.45}
\end{align}
where 
\begin{align}
\mathcal{R}(r)&=\mathcal{K}(r)^2-\Delta_Q\left[r^2+(\mathcal{l}-a\mathcal{e})^2+\mathcal{q}\right]~,\label{1.46}\\
\mathcal{U}(\theta)&=\mathcal{q}-\cos^2\theta\left[a^2(1-\mathcal{e}^2)+\frac{\mathcal{l}^2}{\sin^2\theta}\right]~.\label{1.47}
\end{align}
From eq.(\ref{1.44}), we can obtain the radial trajectory equation in terms of the new modified variable $y$ upto $\mathcal{O}(y)$ as follows
\begin{equation}\label{1.47a}
\begin{split}
(r_+^2+a^2\cos^2\theta_++2r_+y)\frac{dy}{d\tau}&=-\sqrt{\mathcal{P}_0^2-\mathcal{P}_1y+\mathcal{O}(y^2)}~.
\end{split}
\end{equation}
Using the above relation, we can finally write down the form of $\frac{d\tau}{dy}$ up to $\mathcal{O}(y)$ as
\begin{equation}\label{1.48}
\begin{split}
\frac{d\tau}{dy}&\simeq-\frac{\rho_+^2}{\mathcal{P}_0}\left(1+\frac{\mathcal{P}_1y}{2\mathcal{P}_0^2}+\frac{2r_+y}{\rho_+^2}\right)
\end{split}
\end{equation}
where $\rho_+^2=r_+^2+a^2\cos^2\theta_+$, $\mathcal{P}_0=(r_+^2+a^2)(\mathcal{e}-\tilde{\Omega}\mathcal{l}),$ and $\mathcal{P}_1=-4r_+\mathcal{e}\mathcal{P}_0+\Delta'_{Q+}(r_+^2+(\mathcal{l}-a\mathcal{e})^2+\mathcal{q})$. It is easy to solve eq.(\ref{1.48}) and obtain the form of $\tau$ in terms of the spatial variable $y$ as follows (up to $\mathcal{O}(y)$)
\begin{equation}\label{1.49}
\tau(y)\simeq-\frac{\rho_+^2y}{\mathcal{P}_0}.
\end{equation}
Now the energy measured from the viewpoint of the corotating frame (dragged with an angular velocity $\tilde{\Omega}$) reads $\tilde{\mathcal{e}}=(\mathcal{e}-\tilde{\Omega}\mathcal{l})$ implying that the constant $\mathcal{P_0}$ as well as the leading order term is eq.(\ref{1.49}) is proportional to $\tilde{\mathcal{e}}$. Now using eq.(\ref{1.41}) along with eq.(\ref{1.44}), we obtain the form of the time $t$ up to $\mathcal{O}(y)$ as 
\begin{equation}\label{1.50}
t(y)\simeq-\frac{1}{2\kappa}\ln y-\delta y
\end{equation}
where the constant $\delta$ is given as 
\begin{equation}\label{1.51}
\begin{split}
\delta=&\left(\frac{\mathcal{P}_1}{2\mathcal{P}_0^2}-\frac{1}{2\kappa (r_+^2+a^2)}+\frac{4r_+}{r_+^2+a^2}\left(1+\frac{\tilde{\Omega}\mathcal{l}}{2\tilde{e}}\right)\right)\frac{1}{2\kappa}-\frac{\tilde{\Omega}}{\tilde{e}}(a\mathcal{e}s_+^2-\mathcal{l})
\end{split}
\end{equation}
with $s_+=\sin\theta_+$. We can now obtain the form of $\phi$ up to $\mathcal{O}(y)$ using eq.(s)(\ref{1.42},\ref{1.44}) as follows
\begin{equation}\label{1.52}
\begin{split}
\phi\simeq&-\frac{\tilde{\Omega}}{2\kappa}\ln y+\left(a\mathcal{e}-\frac{\mathcal{l}}{s_+^2}\right)\frac{y}{\mathcal{P}_0}-\frac{\tilde{\Omega}y}{2\kappa}\left(\frac{\mathcal{P}_1}{2\mathcal{P}_0^2}-\frac{1}{2(r_+2+a^2)\kappa}+\frac{2r_+}{r_+^2+a^2}\frac{\tilde{\mathcal{e}}+\tilde{\Omega}\mathcal{l}}{\tilde{e}}\right)~.
\end{split}
\end{equation}
Using eq.(\ref{1.9}), we can redefine the form of $\tilde{\phi}$ up to $\mathcal{O}(y)$ as follows
\begin{equation}\label{1.53}
\tilde{\phi}=\phi-\tilde{\Omega}t=\lambda y
\end{equation}
where the constant $\lambda$ is given by
\begin{equation}\label{1.54}
\lambda=-\frac{1}{(r_+^2+a^2)\tilde{\mathcal{e}}}(a\mathcal{e}s_+^2-\mathcal{l})\left(\tilde{\Omega}a-\frac{1}{s_+^2}\right)+\frac{\tilde{\Omega}r_+}{\kappa(r_+^2+a^2)}~.
\end{equation}
It is important to note that we have kept $\mathcal{O}(y^2)$ order terms in the expansion of the $\Delta_Q$ function and we have kept up to $\mathcal{O}(y)$ terms in the trajectory equations. In principle, one can calculate the transition probability by keeping $\mathcal{O}(y^2)$ terms in the trajectory equations but our main aim in this paper is to investigate the effect of the near horizon correction in the transition probability. For that, we have also kept $\mathcal{O}(y)$ corrections in the field modes in eq.(\ref{1.37}). The reason for keeping $\mathcal{O}(y^2)$ terms in the expansion of $\Delta_Q$ in eq.(\ref{1.14}) is that it would give a $\mathcal{O}(y)$ correction term in the field modes. We are now in a position to compute the analytical form of the excitation probability.
%higher order correction terms in the linear order forms of the trajectory equations and ultimately calculate the transition probability. Here are no controversies present in terms of order calculation as the field-normal mode solution is also kept up to $\mathcal{O}(y)$.}
\section{The transition probability}\label{Section4}
In this case, we are considering a two-level atom freely falling into the outer horizon of a Kerr-Newman black hole and its emitting photons as observed from the viewpoint of an observer corotating with an angular velocity $\tilde{\Omega}$. If the atomic transition frequency is $\omega$ and the ground state and excited states of the two-level atom are defined as $|g\rangle$ and $|e\rangle$ then we can write the atom field interaction Hamiltonian considering only the coupling with a single scalar field as
\begin{equation}\label{1.55}
\hat{H}_I(\tau)=\hbar\mathcal{G}\left(\hat{b}_{\tilde{\nu}}\psi_{\tilde{\nu}lm}(\bm{r}(\tau),t(\tau))+h.c.\right)(|g\rangle\langle e|e^{-i\omega\tau}+h.c.)
\end{equation}
where $\mathcal{G}$ is the atom-filed coupling constant, $\hat{b}_{\tilde{\nu}}$ is the annihilation operator of the scalar field mode. Here the scalar photon is being emitted simultaneously with the excitation of the two-level atom. Hence, we can write down the excitation probability as
\begin{equation}\label{1.56}
\begin{split}
P_{exc}=&\frac{1}{\hbar^2}\left|\int d\tau \left<1_{\tilde{\nu}},e\right|\hat{H}_I(\tau)\left|0_{\tilde{\nu}},g\right>\right|^2\\
=&\frac{1}{\hbar^2}\left|\int_{0}^{y_f} dy \frac{d\tau}{dy}\left<1_{\tilde{\nu}},e\right|\hat{H}_I[\tau(y)]\left|0_{\tilde{\nu}},g\right>\right|^2\\
=&\mathcal{G}^2\left|\int_{0}^{y_f} dy \frac{(-\rho_+^2)}{\mathcal{P}_0}\left[1+\frac{\mathcal{P}_1y}{2\mathcal{P}_0^2}+\frac{2r_+y}{\rho_+^2}\right]\psi_{\tilde{\nu}lm}^*e^{i\omega\tau}\right|^2
\end{split}
\end{equation}
where we have substituted the form of $\frac{d\tau}{dy}$ up to $\mathcal{O}(y)$ from eq.(\ref{1.48}) in the last line of the above equation and $y_f$ denotes the final limit of the integration. We can finally write down the form of the excitation probability as follows
\begin{equation}\label{1.56a}
\begin{split}
P_{exc}\simeq&\frac{\mathcal{G}^2\rho_+^4}{\mathcal{P}_0^2}\left|\int_{0}^{y_f} dy \left[1+\frac{2r_+y}{\rho_+^2}+\frac{\mathcal{P}_1y}{2\mathcal{P}_0^2}\right]\psi_{\tilde{\nu}lm}^*(y)e^{i\omega\tau}\right|^2.
\end{split}
\end{equation}
In the last line of eq.(\ref{1.56}), we observe that the only contributions in the form of the excitation probability come due to the existence of the counter-rotating terms in the interaction Hamiltonian in eq.(\ref{1.55}). We shall now substitute the form of $\tau$ from eq.(\ref{1.48}) and $\psi^*_{\tilde{\nu}lm}(\bm{r}(y),t(y))$ using eq.(s)(\ref{1.37},\ref{1.49},\ref{1.50}) in the above equation. Keeping terms up to $\mathcal{O}(y)$, we can rewrite the form of the excitation probability as follows
\begin{equation}\label{1.57}
\begin{split}
&P_{exc}=\frac{\mathcal{G}^2\rho_+^4|\Theta_+|^2}{\mathcal{P}_0^2}\left\lvert\int_0^{y_f} dy \biggr[1+\biggr[\frac{2r_+}{\rho_+^2}+\frac{\mathcal{P}_1}{2\mathcal{P}_0^2}+\biggr[\frac{\kappa^2\mathcal{B}}{\kappa^2+\tilde{\nu}^2}7-\frac{1}{2\Delta'_{Q+}}\biggr]\biggr]y+\frac{i\tilde{\nu}\mathcal{B}\kappa}{\kappa^2+\tilde{\nu}^2}y\biggr]y^{-\frac{i\tilde{\nu}}{2\kappa}}e^{-\frac{i\tilde{\nu}}{2\kappa}\ln y}e^{-i\left[m\lambda+\tilde{\nu}\delta+\frac{\rho_+^2\omega}{\mathcal{P}_0}\right]y}\right\rvert^2~.
\end{split}
\end{equation}
We can rewrite the above form of the excitation probability in a much simpler form as follows
\begin{equation}\label{1.58}
\begin{split}
P_{exc}=\mathcal{G}^2\xi^2\left|\int_0^{y_f}dy\left(1+\mathcal{D}y+i\mathcal{E}y\right)y^{-\frac{i\tilde{\nu}}{\kappa}}e^{-i\chi y}\right|^2
\end{split}
\end{equation}
where we have defined $\xi\equiv\frac{\rho_+^2|\Theta_+|}{\mathcal{P}_0}$, $\chi\equiv m\lambda+\tilde{\nu}\delta+\frac{\rho_+^2\omega}{\mathcal{P}_0}$, $\mathcal{D}\equiv\frac{2r_+}{\rho_+^2}+\frac{\mathcal{P}_1}{2\mathcal{P}_0^2}+\frac{\kappa^2\mathcal{B}}{\kappa^2+\tilde{\nu}^2}-\frac{1}{2\Delta'_{Q+}}$, and $\mathcal{E}\equiv\frac{\tilde{\nu}\mathcal{B}\kappa}{\kappa^2+\tilde{\nu}^2}$. It is now important to take a more detailed look into the two oscillatory functions present in the integral.  We shall inspect the two functions $e^{-i\chi y}$ and $y^{-\frac{i\tilde{\nu}}{\kappa}}$. In the $\chi\gg\frac{\tilde{\nu}}{\kappa}$ limit, it is possible to extend the integration limit from $y_f$ to infinity. For the fall of the atom, the geometric optics approximation states $\omega\gg\nu$. If the above condition is satisfied, it is easy to check that $\chi\gg\frac{\tilde{\nu}}{\kappa}$. 
If the above condition is satisfied, it is easy to infer that $e^{-i\chi y}$ behaves as a highly oscillatory function with fixed frequency value, whereas the $y^{-\frac{i\tilde{\nu}}{\kappa}}$ has a ``Russian Doll like behaviour" as discussed in \cite{Ordonez2} for the Kerr black hole case (along with near horizon approximation). As a result $y^{-\frac{i\tilde{\nu}}{\kappa}}$ becomes slowly varying compared to $e^{-i\chi y}$ for higher values of $y$ (for $y>1$). Hence, for higher values of $y$ the overall contribution to the integral becomes negligible which allows one to extend the upper limit of integration from $y_f$ to infinity.
%It is now more convenient to consider $\chi=60$ and $\frac{\tilde{\nu}}{\kappa}=2$ and plot these two oscillatory functions with respect to $y$. We at first plot the two functions in the regime $y\in[0,1]$ in Fig.(\ref{FP1}). 
%\begin{figure}[ht!]
%\begin{center}
%\includegraphics[scale=0.36]{ImgFinal.jpg}
%\caption{Plot of the real values of the functions $e^{-60iy}$ and $y^{-2i}$ vs $y$ for $y\in[0,1]$ and a zoomed in plot in the range $y\in[0.05,0.15]$.\label{FP1}}
%\end{center}
%\end{figure}
%We can see from Fig.(\ref{FP1}), we find that the change in the exponential function ($e^{-60iy}$) is comparable to the change in the $y^{-2 i}$ in the range $y\in[0.05,0.15]$. If we now extend the plot for higher values of $y$, we can find that the change in the $y^{-2i}$ function is much slower with respect to the exponential function for higher values of $y$ due to its ``Russian doll-like behaviour"\cite{Ordonez1} as can be seen from Fig.(\ref{FP2}). Hence, for higher values of $y>1$, the contribution to the integral becomes negligible and we can extend the integral limit from $y_f$ to infinity.
%\begin{figure}[ht!]
%\begin{center}
%\includegraphics[scale=0.36]{ImgFinal2.jpg}
%\caption{Plot of the real values of the functions $e^{-60iy}$ and $y^{-2i}$ vs $y$ for $y\in[1,4]$.\label{FP2}}
%\end{center}
%\end{figure}
Following this argument, it is possible to extend the integration limit in eq.(\ref{1.58}) from $y_f$ to $\infty$ for the first part of the integral in eq.(\ref{1.58}) and we can recast eq.(\ref{1.58}) as follows
\begin{equation}\label{1.58a}
\begin{split}
P_{exc}=&\mathcal{G}^2\xi^2\Bigr|\int_0^{\infty}dy~y^{-\frac{i\tilde{\nu}}{\kappa}}e^{-i\chi y}+\left(\mathcal{D}+i\mathcal{E}\right)\int_0^{y_f}dy~y^{1-\frac{i\tilde{\nu}}{\kappa}}e^{-i\chi y}\Bigr|^2~.
\end{split}
\end{equation}
\noindent We would like to make a further comment here. It is to be noted that for the first term in the integrand in eq.(\ref{1.58}), the integration limit can always be extended from $y_f$ to infinity following the argument given above. However, for the second and the third terms, keeping the limit of the integration finite would give rise to incomplete gamma functions as we shall observe in the following analysis.
We start by evaluating the first part of the integral in eq.(\ref{1.58a}). The first integral reads
\begin{equation}\label{1.58p1}
\begin{split}
\int_0^\infty dy~y^{-\frac{i\tilde{\nu}}{\kappa}}e^{-i\chi y}=-\frac{\tilde{\nu}}{\kappa\chi}e^{-\frac{\pi\tilde{\nu}}{2\kappa}}\chi^{\frac{i\tilde{\nu}}{\kappa}}\Gamma\left(-\frac{i\tilde{\nu}}{\kappa}\right)~.
\end{split}
\end{equation}
For the finite integral given in eq.(\ref{1.58a}), we can evaluate the integral to get the following result
\begin{equation}\label{1.58p2}
\int_0^{y_f}dy~y^{1-\frac{i\tilde{\nu}}{\kappa}}e^{-i\chi y}=-\frac{1}{\chi^2}e^{-\frac{\pi\tilde{\nu}}{2\kappa}}\chi^{\frac{i\tilde{\nu}}{\kappa}}\gamma\left(2-\frac{i\tilde{\nu}}{\kappa},iy_f\chi\right)
\end{equation}
where $\gamma\left(2-\frac{i\tilde{\nu}}{\kappa},iy_f\chi\right)$ is the lower incomplete gamma function. Now, the lower incomplete gamma function can also be represented in terms of the upper incomplete gamma function as
\begin{equation}\label{1.58inc}
\gamma\left(2-\frac{i\tilde{\nu}}{\kappa},iy_f\chi\right)=\Gamma\left(2-\frac{i\tilde{\nu}}{\kappa}\right)-\Gamma\left(2-\frac{i\tilde{\nu}}{\kappa},iy_f\chi\right)~.
\end{equation}
\noindent Combining eq.(\ref{1.58p1}) along with eq.(\ref{1.58p2}), we can write down the expression for the complete integral in eq.(\ref{1.58a}) as follows
\begin{equation}\label{1.58total_integral}
\begin{split}
&\int_0^\infty dy~y^{-\frac{i\tilde{\nu}}{\kappa}}e^{-i\chi y}+(\mathcal{D}+i\mathcal{E})\int_0^{y_f}dy~y^{1-\frac{i\tilde{\nu}}{\kappa}}e^{-i\chi y}=-\frac{\tilde{\nu}e^{-\frac{\pi\tilde{\nu}}{2\kappa}}}{\kappa\chi^{1-\frac{i\tilde{\nu}}{\kappa}}}\left[\Gamma\left[-\frac{i\tilde{\nu}}{\kappa}\right]+\frac{\kappa}{\tilde{\nu}}\left[\frac{\mathcal{D}}{\chi}+\frac{i\mathcal{E}}{\chi}\right]\gamma\left(2-\frac{i\tilde{\nu}}{\kappa},iy_f\chi\right)\right]
\end{split}
\end{equation}
and the complex conjugate of the integral given in the above equation reads
\begin{equation}\label{1.58total_conjugate}
\begin{split}
&\int_0^\infty dy~y^{\frac{i\tilde{\nu}}{\kappa}}e^{i\chi y}+(\mathcal{D}-i\mathcal{E})\int_0^{y_f}dy~y^{1+\frac{i\tilde{\nu}}{\kappa}}e^{i\chi y}=-\frac{\tilde{\nu}e^{-\frac{\pi\tilde{\nu}}{2\kappa}}}{\kappa\chi^{1+\frac{i\tilde{\nu}}{\kappa}}}\left[\Gamma\left[\frac{i\tilde{\nu}}{\kappa}\right]+\frac{\kappa}{\tilde{\nu}}\left[\frac{\mathcal{D}}{\chi}-\frac{i\mathcal{E}}{\chi}\right]\gamma\left(2+\frac{i\tilde{\nu}}{\kappa},-iy_f\chi\right)\right].
\end{split}
\end{equation}
Using eq.(s)(\ref{1.58total_integral},\ref{1.58total_conjugate}), we can write down the excitation probability in eq.(\ref{1.58a}) as follows
\begin{equation}\label{1.58Probability}
\begin{split}
P_{exc}&=\frac{2\pi\mathcal{G}^2\xi^2\tilde{\nu}}{\kappa\chi^2}\frac{1}{e^{\frac{2\pi\tilde{\nu}}{\kappa}}-1}+\frac{\mathcal{G}\xi}{\chi}\sqrt{\frac{2\pi\mathcal{G}^2\xi^2\tilde{\nu}}{\kappa\chi^2e^{\frac{\pi\tilde{\nu}}{\kappa}}}}\biggr[\left(\frac{\mathcal{D}}{\chi}+\frac{i\mathcal{E}}{\chi}\right) \frac{e^{i\text{Arg}\left[\Gamma\left(\frac{i\tilde{\nu}}{\kappa}\right)\right]}}{\sqrt{e^{\frac{2\pi\tilde{\nu}}{\kappa}}-1}}\gamma\left(2-\frac{i\tilde{\nu}}{\kappa},iy_f\chi\right)\\&+\left(\frac{\mathcal{D}}{\chi}-\frac{i\mathcal{E}}{\chi}\right)\frac{e^{-i\text{Arg}\left[\Gamma\left(\frac{i\tilde{\nu}}{\kappa}\right)\right]}}{\sqrt{e^{\frac{2\pi\tilde{\nu}}{\kappa}}-1}} \gamma\left(2+\frac{i\tilde{\nu}}{\kappa},-iy_f\chi\right)\biggr]+\frac{\mathcal{G}^2\xi^2}{\chi^2e^{\frac{\pi\tilde{\nu}}{\kappa}}}\biggr(\frac{\mathcal{D}^2}{\chi^2}+\frac{\mathcal{E}^2}{\chi^2}\biggr)\gamma\left(2-\frac{i\tilde{\nu}}{\kappa},iy_f\chi\right)\gamma\left(2+\frac{i\tilde{\nu}}{\kappa},-iy_f\chi\right)~.
\end{split}
\end{equation}
Eq.(\ref{1.58Probability}) is the main result in our paper. It is important to note that the total probability does not comply with the usual Planck-like behaviour but rather has some deformities introduced due to the occurrence of the lower incomplete gamma functions. It is also important to note that the leading term in eq.(\ref{1.58Probability}) is Planck-like in nature and the extra terms will result in a non-Planckian behaviour in the probability distribution. The non-Planckian nature of the transition probability obtained in eq.(\ref{1.58Probability}) indicates that the apparent equivalence to a thermal distribution with the Hawking temperature $T_H=\frac{\kappa}{2\pi}$ observed in \cite{Ordonez2} gets deformed. We would like to make a further comment here. This is in regard to the principle of equivalence discussed in \cite{Fulling2}. The excitation probability of the atom falling into a Schwarzschild black hole was found to be equal to the excitation probability of the atom when a mirror accelerates with respect to it in flat spacetime. This was regarded as a manifestation of the equivalence principle. In the rotating case, in order to check the equivalence principle, one needs to calculate the excitation probability of the atom when the mirror accelerates (with respect to the atom) in a rotating frame. The analysis would obviously be involved, and one would expect the equivalence principle to hold despite the fact that there are non-Planckian terms in the  excitation probability of the atom in eq.(\ref{1.58Probability}).

\noindent Before going into further mathematical intricacies, we need to focus on the fact that all the non-Planckian terms contribute to the overall transition probability in eq.(\ref{1.58Probability}) only for nonvanishing values of $y_f$. The existence of a very small value of $y_f$ indicates the fact that due to the ``beyond the near horizon" approximation, the finite path traveled by the two-level atom, before freely falling into the event horizon of the black hole, affects the transition spectrum. It can also be claimed that the emitted radiation is no more identical to a black-body spectrum which is an ideal case scenario. The modified transition probability has deviations from the Planckian spectrum as in strong curvature regimes (especially in the vicinity of the event horizon of the black hole) the emission spectrum should be affected by the path traveled by the two-level atom.

\noindent We shall now do a side-by-side comparison with the earlier results present in the literature \cite{Ordonez2}. As our case represents the Kerr-Newman case, we shall be taking the $Q=0$ limit for a side-by-side comparison with the earlier results. In \cite{Ordonez2}, the excitation probability for photon (scalar) emission due to a two-level atom falling into the event horizon of a Kerr black hole was obtained as
\begin{equation}\label{Old_Transition}
P_{exc}^{NH}=\frac{2\pi\mathcal{G}\tilde{\nu}}{\kappa\omega^2}\frac{1}{e^{\frac{2\pi\tilde{\nu}}{\kappa}}-1}~.
\end{equation}
Even in the $Q=0$ limit eq.(\ref{1.58Probability}) is significantly different from the above transition probability. For $\omega\gg \tilde{\nu},m\lambda$, $y_f\rightarrow0$, $Q=0$ and $|\Theta_+|=1$, we get back eq.(\ref{Old_Transition}) from eq.(\ref{1.58Probability}). This comparison shows that eq.(\ref{1.58Probability}) is a more general result capturing the effects of the Kerr geometry outside the event horizon of the black hole. The result also encompasses the amplitude correction introduced via the $|\Theta_+|$ term (coming through $\xi$) due to the $\theta$ part of the covariant Klein-Gordon equation.

\noindent We shall now compute the probability for the de-excitation of the two-level atom from the excited state to the ground state along with a simultaneous absorption of a photon with frequency $\tilde{\nu}$. The initial and the final states for such a process are given as
\begin{align}
|\text{initial}\rangle&=|0_{\tilde{\nu}},e\rangle~,~~|\text{final}\rangle=|1_{\tilde{\nu}},g\rangle~.\label{1.60}
\end{align}
With the redefinition of a constant $\chi'\equiv\frac{\rho_+^2\omega}{\mathcal{P}_0}- m\lambda-\tilde{\nu}\delta$  and using the forms of the initial and final states from eq.(\ref{1.60}), we can calculate the absorption probability as follows
\begin{equation}\label{1.58AbsorptionProbability}
\begin{split}
P_{abs}=&\frac{2\pi\mathcal{G}^2\xi^2\tilde{\nu}}{\kappa\chi^{'2}}\frac{1}{1-e^{\frac{-2\pi\tilde{\nu}}{\kappa}}}-\frac{\mathcal{G}\xi}{\chi'}\sqrt{\frac{2\pi\mathcal{G}^2\xi^2\tilde{\nu}}{\kappa\chi^{'2}e^{-\frac{\pi\tilde{\nu}}{\kappa}}}}\biggr[\left(\frac{\mathcal{D}}{\chi'}+\frac{i\mathcal{E}}{\chi'}\right)\frac{e^{i\text{Arg}\left[\Gamma\left(\frac{i\tilde{\nu}}{\kappa}\right)\right]}}{\sqrt{1-e^{-\frac{2\pi\tilde{\nu}}{\kappa}}}}\gamma\left(2-\frac{i\tilde{\nu}}{\kappa},-iy_f\chi'\right)\\-&\left(\frac{\mathcal{D}}{\chi'}-\frac{i\mathcal{E}}{\chi'}\right)\frac{e^{-i\text{Arg}\left[\Gamma\left(\frac{i\tilde{\nu}}{\kappa}\right)\right]}}{\sqrt{1-e^{\frac{-2\pi\tilde{\nu}}{\kappa}}}} \gamma\left(2+\frac{i\tilde{\nu}}{\kappa},iy_f\chi'\right)\biggr]+\frac{\mathcal{G}^2\xi^2}{\chi^{'2}e^{-\frac{\pi\tilde{\nu}}{\kappa}}}\biggr(\frac{\mathcal{D}^2}{\chi^{'2}}+\frac{\mathcal{E}^2}{\chi^{'2}}\biggr)\gamma\left(2-\frac{i\tilde{\nu}}{\kappa},-iy_f\chi'\right)\gamma\left(2+\frac{i\tilde{\nu}}{\kappa},iy_f\chi'\right)~.
\end{split}
\end{equation}

\section{Plots depicting the nature of the transition probability}\label{Section_Plot}
\begin{figure}[ht!]
\begin{center}
\includegraphics[scale=0.36]{"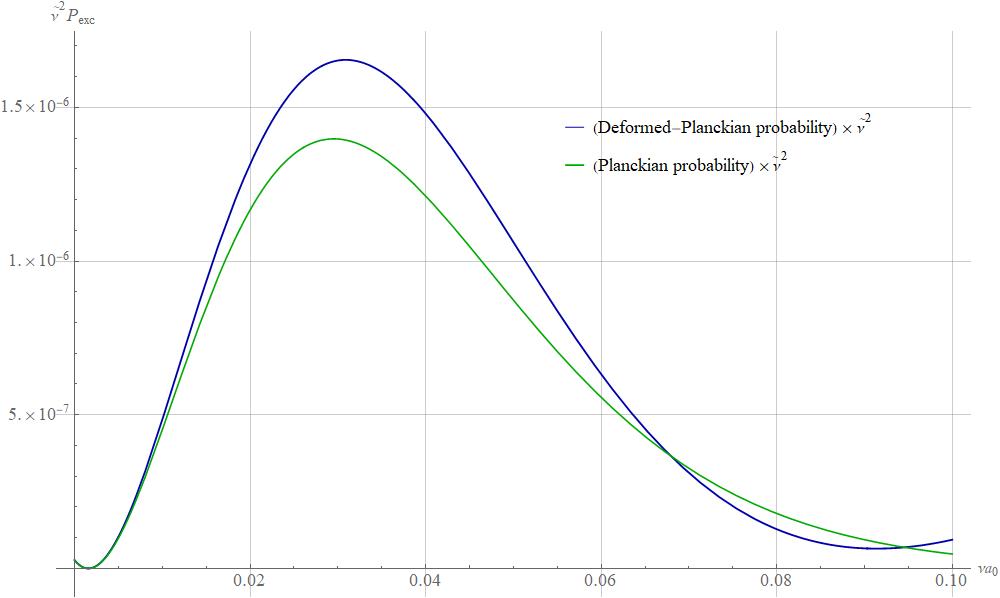"}
\caption{$\tilde{\nu}^2P_{exc}$ vs frequency ($\nu\mathcal{a}_0$ is a dimensionless number) plot for the probability obtained in eq.(\ref{1.58Probability}) times $\tilde{\nu}^2$ vs $\frac{2\pi\mathcal{G}^2\xi^2\tilde{\nu}^3}{\kappa\chi^2}\frac{1}{e^{\frac{2\pi\tilde{\nu}}{\kappa}}-1}$ (usual Planck-like distribution).\label{F2b}}
\end{center}
\end{figure}

In this section, we shall try to understand the nature of the transition probability by plotting various quantities against the frequency of the emitted scalar fields. In order to truly investigate the behaviour of deviation of the probability in eq.(\ref{1.58Probability}), we need to compare it with the usual Planck-like spectrum obtained under a conformal approximation.  In order to plot the probability, we set $M=4 \mathcal{a}_0$, $J=0.4\mathcal{a}_0^2$, $Q=0.35\mathcal{a}_0$, $\sigma_0=0$, $\beta=0.1$, $\mathcal{e}=1$, $m=1$, $\mathcal{l}=1\mathcal{a}_0$, $\theta_+=\frac{\pi}{2}$, and $\Theta_+=1$ (we can always choose the constants $\mathcal{C}_1$ and $\mathcal{C}_2$ in eq.(\ref{1.36}) independently to make $\Theta_+$ unity) with respect to some arbitrary length scale $\mathcal{a}_0$. It is important to remember during the choice of the parameter values that $r_+=M+\sqrt{M^2-a^2-Q^2}$ which provides us with an additional condition that $M\geq \sqrt{a^2+Q^2}$. In this analysis, we have chosen the mass of the scalar field $\sigma_0$ to be zero as we are considering the emission of massless scalar photons. The choice of $\theta_+=\frac{\pi}{2}$ is straightforward as it is most convenient to consider the two-level atom to enter the event horizon parallel to the equatorial plane of the black hole. $\beta$ is a separation constant and can be set to any value possible.  In general, the standard choice is to take the energy per unit mass of the two-level atom ($\mathcal{e}$) to be unity and the quantum numbers $m$ and $l$ to be unity as well. We have maintained the same set of choices while plotting the analytical expressions. For simplicity, we have chosen the atom field coupling constant to be $\mathcal{G}=1 \mathcal{a}_0^{-1}$. As we are in the near horizon (and beyond) limit, it is intuitive to set $y_f$ such that $y_f\ll r_+$ and as a result, we have set the upper limit of integration to be $y_f=\frac{r_+}{10}$~. Choosing comparatively smaller values of $y_f$ makes the deformation of the transition probability much smaller from the Planck-like spectrum. In the plots, we have plotted for $\nu$ values such that the $\tilde{\nu}\ll\omega$ condition is maintained.
\begin{figure}[ht!]
\begin{center}
\includegraphics[scale=0.36]{"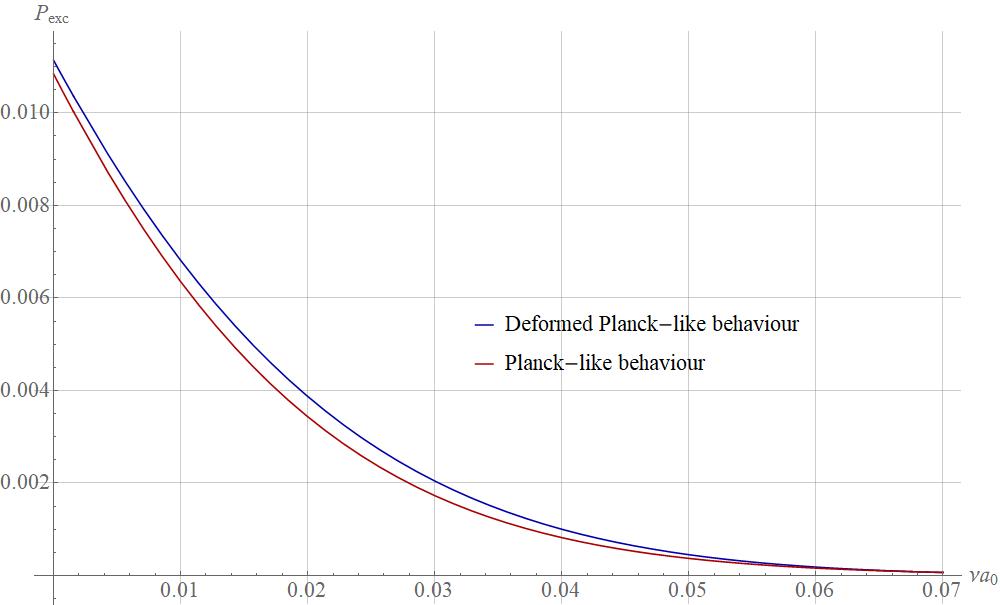"}
\caption{Excitation probability vs frequency ($\nu\mathcal{a}_0$ is a dimensionless number) plot for the probability obtained in eq.(\ref{1.58Probability}) vs $\frac{2\pi\mathcal{G}^2\xi^2\tilde{\nu}}{\kappa\chi^2}\frac{1}{e^{\frac{2\pi\tilde{\nu}}{\kappa}}-1}$ (usual Planck-like distribution).\label{F2a}}
\end{center}
\end{figure}
From Fig.(\ref{F2a}), we can observe that the excitation probability in eq.(\ref{1.58Probability}) has a slight deflection from the usual Planckian probability obtained using a near-horizon analysis. It is important to observe that the deformed Planckian spectrum has a higher area enclosed under the curve denoting a higher chance of detection of the field modes. In Fig.(\ref{F2b}) we have plotted $\tilde{\nu}^2P_{exc}$ vs $\nu\mathcal{a}_0$. From Fig.(\ref{F2b}), one can observe whether the probability is thermal in nature or not. It is straightforward to observe from the plot that the transition probability is thermal for the Planckian probability. It is important to note that for higher values of the field frequencies, the deflection becomes much more significant which may be a cause of the implicit assumption behind the analysis that $\nu\ll\omega$.
\begin{figure}[ht!]
\begin{center}
\includegraphics[scale=0.36]{"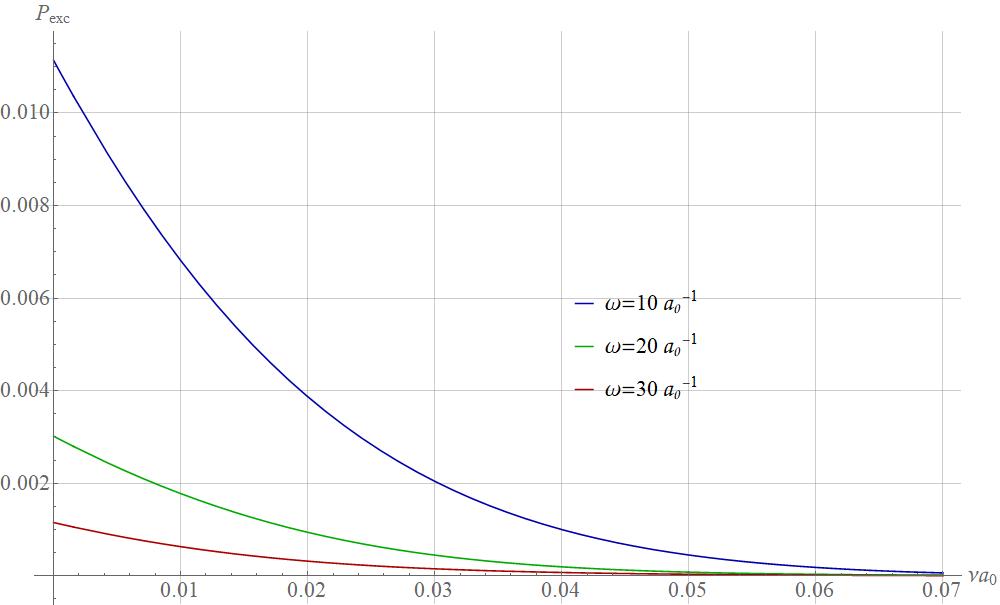"}
\caption{$P_{exc}$ vs frequency ($\nu\mathcal{a}_0$ is a dimensionless number) plot for different values of the detector frequency ($\omega$).\label{F2c}}
\end{center}
\end{figure}
Following the analysis in \cite{BRMajhi}, we shall now plot the excitation probability obtained in eq.(\ref{1.58Probability}) for different values of the detector frequency. From Fig.(\ref{F2c}) we can observe that for higher values of the detector frequency, the area covered under the probability curve reduces significantly indicating a smaller chance of detection of field modes (similar to \cite{BRMajhi}). The above statement can be inferred a priori by looking at eq.(\ref{1.58Probability}). As the $\chi$ term appears in the denominator of all the Planckian as well as non-Planckian terms in the analytical form of the transition probability and $\chi$ is proportional to $\omega$, hence, the value of $P_{exc}$ reduces with increase in the value of the detector frequency $\omega$ leading to a significant reduction in the chance of detecting the individual higher-frequency field modes.

\section{Horizon brightened acceleration radiation entropy}\label{Section6}
We now consider that a stream of two-level atoms is falling into the event horizon of the black hole at a constant rate $k$. If $\Delta\mathcal{N}$ number of particles are falling in a time interval $\Delta t$, we can write
\begin{equation}\label{1.71}
\frac{\Delta\mathcal{N}}{\Delta t}=k.
\end{equation}
We shall now revert to a density matrix approach. We consider that for the $i$-th atom the change in the density matrix is $\delta\varrho_i$. Hence, for $\mathcal{N}$ number of atoms, the change in the density matrix takes the form
\begin{equation}\label{1.72}
\Delta\varrho=\sum\limits_{i=0}^{\Delta\mathcal{N}}\delta\varrho_i=\Delta\mathcal{N}\delta\varrho
\end{equation}
where we have considered that the change in the density matrix is constant for each atom. Using eq.(\ref{1.72}) back in eq.(\ref{1.71}), we get the following relation
\begin{equation}\label{1.73}
\frac{\Delta\varrho}{\Delta t}=k\delta\varrho.	
\end{equation}
The absorption and emission rates can be written as $\Gamma_{exc}=kP_{exc}$ and $\Gamma_{abs}=kP_{abs}$. 
%\begin{equation}\label{1.74}
%\begin{split}
%\dot{\varrho}=&-\frac{\Gamma_{abs}}{2}\left(\varrho b^{\dagger}b+b^\dagger b\varrho-2b\varrho b^\dagger\right)-\frac{\Gamma_{exc}}{2}\left(\varrho bb^{\dagger}+b b^{\dagger}\varrho-2b^{\dagger}\varrho b\right)
%\end{split}
%\end{equation}
In terms of the absorption and emission rates, the rate of change of the $\{n,n\}$-th element of the density matrix is given as (this form can be obtained directly using the Lindblad master equation \cite{Lindblad})
\begin{equation}\label{1.75}
\begin{split}
\dot{\varrho}_{n,n}=&-\Gamma_{abs}\left(n\varrho_{n,n}-(n+1)\varrho_{n+1,n+1}\right)-\Gamma_{exc}\left((n+1)\varrho_{n,n}-n\varrho_{n-1,n-1}\right)~.
\end{split}
\end{equation}
In order to find the steady state single mode solution we need to set $\dot{\varrho}_{n,n}=0$ in the above equation where $\varrho_{n,n}$ can now be replaced by $\varrho_{n,n}^{S.S.}$. Using the above steady-state condition, we can now recast eq.(\ref{1.75}) as follows
\begin{equation}\label{1.75a}
\begin{split}
0=-\Gamma_{abs}\left(n\varrho_{n,n}^{S.S}-(n+1)\varrho_{n+1,n+1}^{S.S}\right)-\Gamma_{exc}\left((n+1)\varrho_{n,n}^{S.S}-n\varrho_{n-1,n-1}^{S.S}\right)~.
\end{split}
\end{equation}
One can now set $n=0$ in the above equation and as a result a $\varrho_{-1,-1}^{S.S}$ term appears which we can ignore as this term is nonexistent. Hence, setting $n=0$ in eq.(\ref{1.75a}) and $\varrho_{-1,-1}^{S.S}=0$, we obtain the following relation
\begin{equation}\label{1.75b}
\begin{split}
0=&\Gamma_{abs}\varrho_{1,1}^{S.S}-\Gamma_{exc}\varrho_{0,0}^{S.S}\\
\implies \varrho_{1,1}^{S.S}=&\frac{\Gamma_{exc}}{\Gamma_{abs}}\varrho_{0,0}^{S.S}~.
\end{split}
\end{equation}
Using the relation obtained in eq.(\ref{1.75b}), in eq.(\ref{1.75a}) for $n=1$, we further obtain $\varrho_{2,2}^{S.S}=\left(\frac{\Gamma_{exc}}{\Gamma_{abs}}\right)^2\varrho_{0,0}^{S.S}$. Following this procedure, we can obtain a recursion relation of the form given by
\begin{equation}\label{1.75c}
\varrho_{n,n}^{S.S}=\left(\frac{\Gamma_{exc}}{\Gamma_{abs}}\right)^n\varrho_{0,0}^{S.S}~.
\end{equation}
Our next aim is to obtain a form of $\varrho_{0,0}^{S.S}$. We use the fact that the trace of any density matrix is 1. Using this property it is straightforward to obtain the following relation
\begin{equation}\label{1.75d}
\begin{split}
&\text{tr}\left[\varrho^{S.S}\right]=\sum_n\varrho^{S.S}_{n,n}=1\\
\implies&\sum_n\left(\frac{\Gamma_{exc}}{\Gamma_{abs}}\right)^n\varrho_{0,0}^{S.S}=1\\
\implies&\frac{1}{1-\frac{\Gamma_{exc}}{\Gamma_{abs}}}\varrho_{0,0}^{S.S}=1\\
\implies &\varrho^{S.S}_{0,0}=1-\frac{\Gamma_{exc}}{\Gamma_{abs}}~.
\end{split}
\end{equation}

\noindent Substituting the analytical form of $\varrho_{0,0}^{S.S}$ from eq.(\ref{1.75d}) in eq.(\ref{1.75c}) we obtain the form of the steady-state single-mode solution as
\begin{equation}\label{1.76}
\varrho_{n,n}^{S.S}=\left(\frac{\Gamma_{exc}}{\Gamma_{abs}}\right)^n\left(1-\frac{\Gamma_{exc}}{\Gamma_{abs}}\right)~.
\end{equation}
In principle if we make use of the excitation and absorption probabilities from eq.(s)(\ref{1.58Probability},\ref{1.58AbsorptionProbability})  to compute the Boltzmann factor ($\Gamma_{exc}/\Gamma_{abs}$), it is evident that the ratio $\Gamma_{exc}/\Gamma_{abs}\neq e^{-\frac{2\pi\tilde{\nu}}{\kappa}}$. This implies that the Boltzmann factor is not thermal in nature. It is also not possible analytically to compute the von-Neuman entropy using eq.(s)(\ref{1.58Probability},\ref{1.58AbsorptionProbability}) due to the complicated analytical forms of the excitation and the absorption probabilities. In order to proceed further, we shall make an approximation where we can extend the integration limit $y_f$ to infinity for the finite integral in eq.(\ref{1.58a}).
Defining a new variable of the form $\chi y\equiv\alpha$, we can rewrite the form of the excitation probability in eq.(\ref{1.58a}) as
\begin{equation}\label{1.59}
\begin{split}
P_{exc}&=\frac{\mathcal{G}^2\xi^2}{\chi^2}\left|\int_0^{\infty}d\alpha\left(1+\frac{\mathcal{D}}{\chi}\alpha+i\frac{\mathcal{E}}{\chi}\alpha\right)\alpha^{-\frac{i\tilde{\nu}}{\kappa}}e^{-i\alpha}\right|^2\\
&=\frac{2\pi\mathcal{G}^2\xi^2\tilde{\nu}}{\kappa^3 \chi^4}(\kappa^2(\mathcal{D}^2+(\mathcal{E}+\chi)^2)+(\mathcal{D}^2+\mathcal{E}^2)\tilde{\nu}^2-2\kappa\mathcal{D}\chi\tilde{\nu})\frac{1}{e^{\frac{2\pi\tilde{\nu}}{\kappa}}-1}~.
\end{split}
\end{equation}
The absorption probability under the same approximation takes the form 
\begin{equation}\label{1.61}
\begin{split}
P_{abs}&=\frac{2\pi\mathcal{G}^2\xi^2\tilde{\nu}}{\kappa^3\chi'^4}(\kappa^2(\mathcal{D}^2+(\mathcal{E}-\chi')^2)+(\mathcal{D}^2+\mathcal{E}^2)\tilde{\nu}^2+2\kappa\mathcal{D}\chi'\tilde{\nu})\frac{1}{1-e^{-\frac{2\pi\tilde{\nu}}{\kappa}}}~.
\end{split}
\end{equation}
We shall now make use of the approximate forms of the probabilities from eq.(s)(\ref{1.59},\ref{1.61}). In order to compute the Horizon Brightened acceleration radiation entropy, we consider $\omega\gg\tilde{\nu},m\lambda$. Following this assumption we find $\chi=\chi'\simeq\frac{\rho_+^2\omega}{\mathcal{P}_0}$ and $\chi>\mathcal{E}$. As additional assumptions, we consider $\chi>\mathcal{D}$ and $\kappa>\tilde{\nu}$. Using the above assumptions, we can write down the Boltzmann factor as 
\begin{equation}\label{1.77}
\begin{split}
\frac{\Gamma_{exc}}{\Gamma_{abs}}=\frac{P_{exc}}{P_{abs}}&=\left(1+\frac{4\mathcal{E}}{\chi}-\frac{4\mathcal{D}\tilde{\nu}}{\kappa \chi}\right)e^{-\frac{2\pi\tilde{\nu}}{\kappa}}\\
&\simeq\left(1+\frac{4\mathcal{B}\tilde{\nu}}{\kappa\chi}-\frac{4\mathcal{D}\tilde{\nu}}{\kappa \chi}\right)e^{-\frac{2\pi\tilde{\nu}}{\kappa}}~.
\end{split}
\end{equation} 
Despite the occurrence of the exponential term, it is evident that the Boltzmann factor is no more thermal.
The von-Neumann entropy for the system reads
\begin{equation}\label{1.78}
S_{\varrho}=-\text{Tr}[\varrho\ln\varrho]=-\sum\limits_{n,\nu,l,m}\varrho_{n,n}\ln\varrho.
\end{equation}
The rate of change of the von Neumann entropy in eq.(\ref{1.78}) can be computed as follows (with the Boltzmann constant set to $1$)
\begin{equation}\label{1.79}
\dot{S}_{\varrho}=-\frac{d}{dt}\left(\text{Tr}[\varrho\ln\varrho]\right)\simeq-\sum\limits_{n,\nu,l,m}\dot{\varrho}_{n,n}\ln\varrho_{n,n}^{S.S.}.
\end{equation}
Now for the ${n,n}$ element of the density matrix with the set of quantum numbers $s=\tilde{\nu,l,m}$, we get the following relation
\begin{equation}\label{1.80}
\sum\limits_{n}\dot{\varrho}_{n,n}n=\dot{\bar{n}}_{s}~.
\end{equation}
If we now sum $\dot{\bar{n}}_{s}\tilde{\nu}$ over all the quantum numbers, we obtain
\begin{equation}\label{1.81}
\begin{split}
\sum\limits_{s=\{\nu,l,m\}}\dot{\bar{n}}_{s}\tilde{\nu}&=\sum\limits_{s=\{\nu,l,m\}}\dot{\bar{n}}_{s}(\nu-m\tilde{\Omega})=\dot{E}_p-\tilde{\Omega}\dot{J}_p~.
\end{split}
\end{equation} 
In eq.(\ref{1.81}) $\dot{E}_p$ is the rate of change in the total energy of the black hole and $\dot{J}_p$ is the rate of change of the axial angular momentum of the black hole (defined about the rotational axis of the black hole). The extra term in eq.(\ref{1.81}) is generated due to the angular momentum of the black hole. Using the definition of $a=\frac{J}{M}$, we can rewrite eq.(\ref{1.81}) as follows
\begin{equation}\label{1.82}
\sum\limits_{s=\{\nu,l,m\}}\dot{\bar{n}}_{s}\tilde{\nu}=\dot{m}_p-\frac{a^2}{r_+^2+a^2}\dot{m}_p=\frac{\dot{m}_pr_+^2}{r_+^2+a^2}
\end{equation}
with $\dot{m}_p$ denoting the rate of change of the mass of the black hole due to emitting photons. In order to express the rate of change in the von Neumann entropy in terms of the area of the black hole, we compute the area of the Kerr-Newman black hole at $r=r_+$ and some fixed time as
\begin{equation}\label{1.83}
A=\int\sqrt{g_{\theta\theta}g_{\phi\phi}}d\theta d\phi=4\pi (r_+^2+a^2)~.
\end{equation}
The rate of change of the area of the black hole due to the simultaneous emission of the photons can be directly related to the rate of change of the mass of the black hole due to emitting photons via the following relation
\begin{equation}\label{1.84}
\dot{A}_p=\frac{16\pi r_+^2}{\Delta'_{Q+}}\dot{m}_p~.
\end{equation}
Substituting eq.(\ref{1.76}) into eq.(\ref{1.79}) and using eq.(s)(\ref{1.80}-\ref{1.84}), we obtain the following form for the rate of change of the von Neumann entropy
\begin{equation}\label{1.85}
\begin{split}
\dot{S}_{\varrho}&=-\sum\limits_{n,\nu,l,m}n\dot{\varrho}_{n,n}\left(\ln\left[1-\frac{4\tilde{\nu}}{\kappa\chi}\left(\mathcal{D}-\mathcal{B}\right)\right]-\frac{2\pi\tilde{\nu}}{\kappa}\right)\\
&=\frac{\dot{m}_pr_+^2}{r_+^2+a^2}\frac{2\pi}{\kappa}+\frac{4\dot{m}_pr_+^2}{\kappa\chi(r_+^2+a^2)}\left(\mathcal{D}-\mathcal{B}\right)\\
\implies \dot{S}_{\varrho}&\simeq\frac{\dot{A}_p}{4}+\frac{\dot{A}_p\mathcal{P}_0}{2\pi\rho_+^2\omega}\left(\frac{2r_+}{\rho_+^2}+\frac{\mathcal{P}_1}{2\mathcal{P}_0^2}-\frac{1}{2\Delta'_{Q+}}\right)~.
\end{split}
\end{equation}
Eq.(\ref{1.85}) is one of the most important results in our work, the first term of the above result can be termed as the ``$\dot{A}/4$ law" in the context of HBAR entropy. It shows that because of the consideration of the beyond near horizon approximation the the rate of change of the HBAR entropy picks up an extra correction term apart from the usual total time derivative of the ``\textit{area divided by four}" form factor \cite{Ordonez4}. 

\noindent Eq.(\ref{1.85}) is very interesting when compared to the result obtained in \cite{Ordonez4}. We start our comparison by looking at the Boltzmann factor obtained in eq.(\ref{1.77}). For a calculation using the near horizon contribution of the black hole, the Boltzmann factor reads \cite{Ordonez4}
\begin{equation}\label{1.86}
\frac{\Gamma_{exc}}{\Gamma_{abs}}=e^{-\frac{2\pi\tilde{\nu}}{\kappa}}~.
\end{equation}
The above factor has a similar form (as discussed in \cite{Ordonez4}) to that of a thermal state with a detailed balance Boltzmann factor
$\Gamma_{exc}/\Gamma_{abs}=e^{-\frac{\tilde{\nu}}{T}}$, $(k_B=1)$. This implies that $T=\frac{\kappa}{2\pi}$, which can be easily seen to be the Hawking temperature ($T_H$) of the black hole. From eq.(\ref{1.77}),  the Boltzmann factor obtained in our current beyond near horizon analysis is given by eq.(\ref{1.77}). Therefore, the temperature obtained in this scenario is equal to (where $\mathcal{D},\mathcal{B}<\chi$) 
\begin{equation}\label{1.87}
T\simeq T_H+\frac{\kappa}{\pi^2\chi}(\mathcal{B}-\mathcal{D})~.
\end{equation}
Eq.(\ref{1.87}) implies that thermality is lost which was present earlier in \cite{Ordonez4}. This behaviour is also reflected in the rate of change of the von Neumann entropy obtained in eq.(\ref{1.85}). The result shows that the entropy does not follow the same behaviour as the Bekenstein-Hawking entropy, rather the new correction term implies that the thermality of the state is now lost due to contributions coming from the beyond-near horizon approximation. There is also a more involved physical aspect of this result. Our result implies that the thermal nature prevails more and more when the atom is very near the event horizon of the black hole. When the atom is a little bit away from the event horizon, the overall nature is not purely thermal rather it approaches a  thermal nature with a decreasing initial distance from the event horizon of the black hole.
\section{Conclusion}
In this work, we have considered a two-level atom freely falling into the event horizon of a rotating and charged black hole. In our analysis, we have considered the event horizon of the black hole to be covered by a perfectly reflecting mirror which shields the infalling atoms from interacting with the outgoing Hawking radiation. This particular setup is identical to that of a Boulware-like vacuum \cite{Boulware}. In order to uniquely define the Boulware vacuum state, one also needs to put a perfectly reflecting and corotating mirror inside the speed of light surface. Our analysis captures the essence of going beyond the near-horizon approximation. In the case of a near horizon analysis, there exists an underlying conformal symmetry. In the beyond near horizon analysis, the underlying conformal symmetry breaks down and the excitation probability no longer retains an overall Planck-like factor as can be seen from eq.(\ref{1.58Probability}). We instead now obtain several terms involving lower incomplete gamma function which leads to a deformed Planck-like behaviour of the excitation probability as can be seen from Fig.(s)(\ref{F2a},\ref{F2b}). In this sense, our analysis extends the main result in \cite{Ordonez2} and supports the statement that the overall Planck-like factor comes due to an underlying conformal symmetry only.  The deformation from the Planckian nature is due to the integration limit $y_f$ being a finite value. It encompasses the effect of the path traveled by the two-level atom before entering the event horizon of the black hole. The deviation of the spectrum from the black body spectrum owes its origin to the curvature of the rotating spacetime affecting the physics significantly when one goes beyond the near horizon approximation. We also have done a side-by-side comparison between the transition probabilities of our paper with \cite{Ordonez2} in the Kerr black hole case. We have then plotted the excitation probability for different values of the detector frequency and from Fig.(\ref{F2c}), it is straightforward to observe that for higher values of the atomic transition frequency, there is a drop in the probability of detecting scalar field-modes.
Finally, we have computed the von-Neumann entropy which is also termed as the horizon brightened acceleration radiation entropy \cite{Fulling2}. We observe that in the analytical form of the rate of change of the von Neumann entropy, apart from the total time derivative of the area divided by four terms, there exist extra correction terms. These correction terms are generated only due to the consideration of the beyond-near horizon approximation. 
\section*{Data availability statement}
The manuscript has no data associated with it.
\section*{Acknowledgement}
We thank the referee for useful and constructive comments which have helped us to improve the paper substantially. 

\end{document}